\documentstyle[12pt]{article}
\newlength{\bredde}
\def\slash#1{\settowidth{\bredde}{$#1$}\ifmmode\,\raisebox{.15ex}{/}
\hspace*{-\bredde} #1\else$\,\raisebox{.15ex}{/}\hspace*{-\bredde} #1$\fi}
\newcommand{\beq}{\begin{equation}}
\newcommand{\eeq}{\end{equation}}

\def\gtwid{\raise.3ex\hbox{$>$\kern-.75em\lower1ex\hbox{$\sim$}}}
\def\ltwid{\raise.3ex\hbox{$<$\kern-.75em\lower1ex\hbox{$\sim$}}}
\begin{document}
\vspace*{1cm}
\topmargin -0.8cm
\oddsidemargin -0.8cm
\evensidemargin -0.8cm
\headheight 0pt
\headsep 0pt
\topskip 9mm
\title{\Large{
Master Equations for Extended Lagrangian BRST Symmetries}}

\vspace{0.5cm}

\author{{\sc Poul H. Damgaard} and {\sc Frank De Jonghe}\thanks{
Aspirant of the N.F.W.O., Belgium. On
leave from the Instituut voor Theoretische Fysica, University of Leuven,
Belgium.}\\
CERN -- Geneva
}
\maketitle
\vfill
\begin{abstract} Starting from the requirement that a Lagrangian field
theory be invariant under both Schwinger-Dyson BRST and
Schwinger-Dyson anti-BRST symmetry, we derive the BRST--anti-BRST
analogue of the Batalin-Vilkovisky formalism. This is done through
standard Lagrangian gauge fixing respecting the extended BRST
symmetry. The solutions of the resulting Master Equation and
the gauge-fixing procedure for the quantum action can be brought
into forms that coincide with those obtained earlier on algebraic
grounds by Batalin, Lavrov and Tyutin.
\end{abstract}
\vfill
\vspace{6.5cm}
\begin{flushleft}
hepth@xxx/9302050 \\
KUL-TF-93/04 \\
CERN--TH-6800/93 \\
February 1993
\end{flushleft}
\vfill
\newpage


Schwinger-Dyson equations are believed to provide a complete
description of any quantum field theory, once appropriate boundary
conditions are imposed. To implement the content of these
Schwinger-Dyson equations as a dynamical principle in Lagrangian
quantization, one needs the symmetry operator that
enforces the equations. This ``Schwinger-Dyson BRST operator",
in the following denoted by $\delta$, was first found in
ref. \cite{us0}. (For some generalizations, see ref.
\cite{us1}).

Recently, it was discovered \cite{us} that demanding a Lagrangian
field theory to be invariant under this Schwinger-Dyson BRST symmetry
in addition to usual BRST symmetries of internal gauge invariances
leads directly to the Batalin-Vilkovisky Lagrangian formalism
\cite{Batalin}. The derivation is done most simply in terms of
a certain Lagrangian collective field formalism, described in
detail in ref. \cite{us1}, but can also be based directly on the
Schwinger-Dyson BRST symmetry itself.

One interesting aspect of these observations is that the
Batalin-Vilkovisky formalism, normally based on a one-to-one
doubling of field variables\footnote{For some excellent
recent reviews of the Batalin-Vilkovisky formalism, see
ref. \cite{review}.} (to each fundamental ``field" $\phi^A$,
including usual ghosts and antighosts whenever needed, one
associates an ``antifield" $\phi^*_A$), actually corresponds to an
intermediate
step of the Lagrangian quantization procedure, in which a certain
set of ghost fields $c^A$ already have been integrated out.
These new ghosts $c^A$ play a fundamental r\^{o}le in the
Schwinger-Dyson BRST transformations \cite{us0}. When the functional
measures for the fundamental fields are flat, the Schwinger-Dyson
BRST algebra reads (our conventions follow those of ref. \cite{us}):
\begin{eqnarray}
\delta \phi^A(x) &=& c^A(x) \cr
\delta c^A(x) &=& 0 \cr
\delta \phi^*_A(x) &=& -\frac{\delta^l S}{\delta\phi^A(x)} ~,
\end{eqnarray}
with $S$ generically indicating the full action. The Ward
Identities following from this symmetry, when implemented in the
Lagrangian path integral, are precisely the most general
Schwinger-Dyson equations. The way to include internal gauge
symmetries of the starting point, the given classical action
$S_0$, is described in ref. \cite{us}. After integrating out
the new ghosts $c^A$, the method is completely equivalent to the
Batalin-Vilkovisky scheme \cite{Batalin}. The antifields $\phi^*_A$
are simply the usual antighosts of the ghosts $c^A$ that enforce
Schwinger-Dyson equations through shifts of the fundamental fields.

Algebraically, the Schwinger-Dyson BRST transformations provide
a symmetry of the full quantum partition function whenever the
quantum action $S_{ext}$ satisfies a Master Equation of the form
\beq
\frac{1}{2}(S_{ext},S_{ext}) = - \frac{\delta^r S_{ext}}
{\delta\phi^A}c^A + i\hbar\Delta S_{ext}~,
\eeq
with the antibracket $(\cdot,\cdot)$ and the operator $\Delta$
coinciding with those of Batalin and Vilkovisky \cite{Batalin}.
It is important to realize that the above Master Equation must be
imposed from the beginning, independently of any possible gauge
fixings. For theories of flat measures (and we shall consider only
those cases here), eq. (2) is a necessary condition that
the final quantum field theory gives rise to the correct
Schwinger-Dyson equations. Gauge-independence of the generating
functional upon any subsequent gauge fixings follows
straightforwardly as a by-product \cite{us}.

A sufficient condition to guarantee that the resulting Ward
Identities of $S_{ext}$ contain all Schwinger-Dyson equations
for gauge-invariant Green functions of the
classical fields is that the solution $S_{ext}$
of eq. (2) can be written
\beq
S_{ext}[\phi,\phi^*,c] = S^{(BV)}[\phi,\phi^*] - \phi^*_Ac^A~,
\eeq
with $S^{(BV)}$ then required to satisfy the quantum Master
Equation of Batalin and Vilkovisky:
\beq
\frac{1}{2}(S^{(BV)},S^{(BV)}) = i\hbar\Delta S^{(BV)}~.
\eeq

Upon integrating out $c^A$, this automatically sets $\phi^*_A
=0$ before gauge fixing. This also, together with the
Batalin-Vilkovisky boundary conditions \cite{Batalin} imposed
on $S^{(BV)}$, guarantees that the final gauge fixing is
achieved by a certain canonical transformation (within the
antibracket).

The resulting Lagrangian formalism corresponds directly to (but
provides a much more powerful machinery than) standard BRST
Lagrangian quantization. In particular, the action $S_{ext}$ itself
can in a certain specific sense \cite{Batalin} be viewed as the
Lagrangian BRST generator.

As formulated above, this seems to leave out the anti-BRST
symmetries that can always be constructed for closed
algebras, given the BRST
symmetries \cite{anti}. However, it was shown already in
ref. \cite{us0} that there also exists a corresponding
{\em Schwinger-Dyson anti-BRST symmetry}. The question then
naturally arises whether there also
exists a corresponding Lagrangian anti-BRST generator. In
particular, can one set up a corresponding quantization
prescription that leads to a path integral invariant under
{\em extended} BRST symmetry, $i.e.$ simultaneous
invariance under both BRST and anti-BRST symmetries? This question
has been answered in the affirmative by Batalin, Lavrov and Tyutin
\cite{Ext} from a completely
algebraic starting point.\footnote{For
some alternative formulations, see also ref. \cite{Gomis}.}
For those who find the standard Batalin-Vilkovisky scheme
(with its doubling of fields and subsequent conditions on
how to remove the antifields) rather formidable, the
extended BRST scheme of Batalin, Lavrov and Tyutin must appear
even more impressive. It requires the introduction of now {\em
three} antifields $\phi^*_{Aa}, a = 1,2$ and $\bar{\phi}^A$,
all of which are finally removed through a certain gauge-fixing
procedure that freezes their values \cite{Ext}.

If the ordinary Batalin-Vilkovisky Lagrangian scheme follows from
imposing the Schwinger-Dyson BRST symmetry on the path integral,
one would expect that the most natural starting point for
deriving a path integral invariant under extended BRST symmetries
would be to impose simultaneously invariance under both the
Schwinger-Dyson BRST {\em and} Schwinger-Dyson
anti-BRST symmetries. Because the algebra corresponding to the
Schwinger-Dyson anti-BRST symmetry \cite{us0} can be formulated
in terms of the same multiplet of fields needed for the
corresponding BRST symmetry, one would a priori not expect new
fields to appear. One purpose of the present letter is to clarify
that issue. Starting from the same collective field technique
employed in ref. \cite{us}, we shall {\em derive} the Master
Equation that gives rise to an action of extended BRST symmetry
in the case of an irreducible closed algebra.

We start by recapitulating the BRST and anti-BRST transformations
for Schwinger-Dyson equations \cite{us0}. We will change notation
and denote the BRST
operator by $\delta_1$ and its anti-BRST companion by $\delta_2$.
As we wish to treat both symmetries on an equal footing, which means
that the ghosts $c^A$ are viewed as $Sp(2)$ partners of the antighosts
$\phi^*_A$, we choose a more symmetric notation in which $c^A$ and
$\phi^*_A$ are considered as the first and second component of
a combined field $\phi^*_{Aa}, a = 1,2$.

In order to see the extended Schwinger-Dyson symmetries
and their consequences for the action, it suffices to consider
a theory of classical action $S[\phi]$ that is free of internal
gauge invariances. Introducing collective
fields $\bar{\phi}^A$ for the $\phi^A$-fields\footnote{The choice
of notation for the collective fields ($\bar{\phi}^A$) differs from
that of ref. \cite{us}. The reason for the present choice will
become obvious below.}, we gauge the shift symmetry of the flat
$\phi^A$-measure through the transformation $\phi^A \to \phi^A
- \bar{\phi}^A$, and integrate over these collective fields
$\bar{\phi}^A$ as well. The fields $\phi^A$ are of arbitrary
Grassmann parity $\epsilon_A$, and we of course require
$\epsilon(\bar{\phi}^A) = \epsilon_A$. There is no distinction
between upper and lower indices on $\phi$. The BRST (and anti-BRST)
versions of the above gauge symmetry are the seeked-for
Schwinger-Dyson BRST (and anti-BRST) symmetries \cite{us0}:
\begin{eqnarray}
\delta_a\phi_A &=& \phi^*_{Aa} \cr
\delta_a\bar{\phi}_A &=& \phi^*_{Aa} \cr
\delta_a\phi^*_{Ab} &=& \varepsilon_{ab}B_A \cr
\delta_a B_A &=& 0~,
\end{eqnarray}
where $\varepsilon_{ab}$ is the $Sp(2)$ invariant tensor
($\varepsilon_{12} = 1, \varepsilon_{ab} = -\varepsilon_{ba}$).

It follows trivially that these transformations are both nilpotent
and satisfy the correct BRST--anti-BRST algebra
\beq
\delta_a\delta_b + \delta_b\delta_a = 0 ~.
\eeq

Using the fact that $gh(\delta_a)= -(-1)^a$ we derive the following
ghost number assignments:
\begin{eqnarray}
   gh(\phi^*_{A a}) & = & - (-1)^a + gh(\phi^A) \cr
   gh(\bar{\phi}^A) & = & gh(\phi^A) \cr
   gh(B^A) & = & gh(\phi^A)~.
\end{eqnarray}
Grassmann parities of the new fields are $\epsilon(\phi^*_{Aa}) =
\epsilon_A + 1$ and $\epsilon(B_A) = \epsilon_A$.

Gauge-fixing the collective fields to zero in a manner manifestly
invariant under the extended Schwinger-Dyson BRST can be achieved
by adding a term
\beq
\frac{1}{4}\varepsilon^{ab}\delta_a\delta_b
[\bar{\phi}_A (-1)^{\epsilon_B}M_{AB}\bar{\phi}_B]
= (-1)^{\epsilon_B+1}\bar{\phi}_AM_{AB}B_B
- \frac{1}{2}\varepsilon^{ab}\phi^*_{Aa}M_{AB}\phi^*_{Bb}~.
\eeq
Here $M_{AB}$ is a constant invertible matrix of symmetry $M_{AB} =
(-1)^{\epsilon_A\cdot\epsilon_B}M_{BA}$. All elements of $M$ between
Grassmann odd and even sectors vanish. $M$ ensures that the term we
add to the action has overall ghost number zero, and is even under
Grassmann parity.\footnote{This matrix $M$
is also required to gauge-fix in a BRST--anti-BRST preserving manner
Grassmann-odd fields to zero. Its presence may be avoided
by the introduction of more fields in the extended BRST multiplet,
but for our purposes the present formulation suffices. $M$ will
drop out in the end.}

Integrating out $\bar{\phi}_A$ and $B_A$ yields the action
\beq
S_{ext} = S[\phi] - \frac{1}{2}\varepsilon^{ab}\phi^*_{Aa}
M_{AB}\phi^*_{Bb}~,
\eeq
which is invariant under the more compact version of the extended
Schwinger-Dyson BRST symmetry
\begin{eqnarray}
\delta_a\phi_A &=& \phi^*_{Aa} \cr
\delta_a\phi^*_{Ab} &=& \varepsilon_{ba}M_{BA}^{-1}
\frac{\delta^l S_{ext}}{\delta\phi_B}
\end{eqnarray}
obtained by substituting for $B_A$ the equation of motion. Ward
Identities of the form $0 = \langle \varepsilon^{ab}
\delta_b\phi^*_{Aa}F[\phi] \rangle$
are the most general Schwinger-Dyson equations
\beq
\langle \frac{\delta^lF}{\delta\phi_A(x)} +
\left(\frac{i}{\hbar}\right)\frac{\delta^lS}{\delta\phi_A(x)}
F[\phi_A] \rangle = 0
\eeq
of the partition function
\beq
{\cal{Z}} = \int [d\phi^A][d\phi^*_{Aa}]
\exp\left[\frac{i}{\hbar}S_{ext}\right]~.
\eeq

Turning the argument around, we can now write down algebraic
equations the extended action $S_{ext}$ must satisfy in order to
be invariant under the extended Schwinger-Dyson BRST symmetry
(10). Inserting the transformations (10) into the conditions
$0 = \delta_a S_{ext}$ is seen to be equivalent
to demanding that $S_{ext}$ must satisfy the following Master
Equations:
\beq
\varepsilon_{ba}\frac{\delta^r S_{ext}}{\delta\phi^*_{Aa}}
M_{BA}^{-1}\frac{\delta^l S_{ext}}
{\delta\phi^B} = \frac{\delta^r S_{ext}}
{\delta\phi^A}\phi^*_{Ab}~.
\eeq

This is just the $Sp(2)$ invariant version of
the classical part of the Master Equation (2). Because of the
derivation in terms of collective fields, we know that the
functional measure (formally) remains invariant.\footnote{The
action and the starting measure are trivially
invariant, and the combination of action and measure
must respect the symmetry even after having integrated out some
of the fields. The measure therefore must remain invariant. This
can of course also be checked explicitly.} We are
throughout assuming the existence of a suitable regularization
scheme that respects the BRST symmetry. If one insists on regulators
that break the symmetry, the problem must be cast in terms of the
quantum Master Equation, to be discussed briefly below.
(Some aspects of
this issue in the usual Batalin-Vilkovisky BRST framework are
discussed in ref. \cite{Troost}.)

Next, we turn to the question of internal gauge symmetries.
In the case of just BRST symmetry, it was found that the inclusion
of gauge symmetries did not fundamentally alter the formalism.
In particular, the Master Equation remains the same. This is as
one would expect, because Schwinger-Dyson equations should be
satisfied independently of particular choices of gauge fixing.
As we have mentioned above, a solution of the kind (3) ensures
that the resulting equations for the original {\em classical}
fields coincide with those of the original action $S$. Introducing
the Batalin-Vilkovisky boundary conditions \cite{Batalin}
implies that the final gauge fixing procedure can be done through
a canonical transformation alone. Different boundary conditions
can be equally valid for the case of closed algebras, but then
the gauge-fixing procedure may require the usual addition of
a BRST-exact term \cite{us}.

Imposing only anti-BRST symmetry can be done in a completely
similar manner, and we shall not elaborate on it.

However, insisting on having {\em extended} BRST--anti-BRST
symmetry brings in a fundamentally new aspect. In order to set
up the gauge-fixing of the extended action $S_{ext}$ satisfying
the Master Equation (13), one must find the equivalent of
adding a gauge-fixing term that is simultaneously invariant
under both the BRST and anti-BRST symmetries. Normally, this
is done by adding a term of the form $\varepsilon^{ab}\delta_a
\delta_b \chi(\phi)$, for a suitable bosonic function $\chi$.
But after having integrated out the auxiliary fields $B_A$,
the operator $\varepsilon^{ab}\delta_a\delta_b$ is no longer
nilpotent, even when acting on the subspace of fields $\phi_A$.
This means that there is no simple way of starting directly
with the Master Equation (13), introducing suitable boundary
conditions, and then finding an appropriate way of gauge
fixing.\footnote{This problem does not arise if one insists only
on preserving either BRST or anti-BRST invariance. A sufficient
gauge-fixing term is then of the form $\delta_a\Psi(\phi)$ for
a certain fermionic function $\Psi$, and the operator $\delta_a$
{\em is} nilpotent when acting only on the subspace of $\phi_A$,
-- which is all that is required.}

This problem is instantly cured if we revert to the formulation
in terms of collective fields $\bar{\phi}_A$. The operator
$\varepsilon^{ab}\delta_a\delta_b$ is now by construction
nilpotent, and we can use it to add BRST--anti-BRST exact
terms to the action without changing the physical content. But of
course this also means that we are restricted to situations
where the collective field formalism can be applied. In the case
of open gauge algebras a consistent formalism that
allows us to merge the collective field BRST symmetry with that
of the internal gauge algebra has not yet been found. Thus by
remaining at the level where the extended BRST symmetry is
still formulated with the help of the collective fields
discussed above (and no more), we
are at present prevented from constructing, for open gauge
algebras, a gauge-fixed action invariant under the extended
BRST symmetries. It is not yet clear whether this implies that
extended BRST symmetries cannot be imposed on such theories.
Introducing more fields may be one way of extending properly
the collective field formalism to that case.

The starting point, then, is that
for a closed irreducible gauge algebra, the existence of both
BRST and anti-BRST transformations \cite{anti} leads us to
consider {\em two} nilpotent transformations induced by the
internal gauge symmetry. As is customary, $\phi_A$ denotes
collectively all classical fields, ghosts and antighosts, and
auxiliary fields. The BRST and anti-BRST transformations for
the internal symmetries can then be written in the general
for $\delta_a\phi_A = \cal{R}_{Aa}$. These transformations
satisfy (6) and $\delta_a\cal{R}_{Aa} = 0$ for both values
of $a$.
To end the cycle, one must introduce suitable Nakanishi-Lautrup
fields, which are now forced to transform non-trivially.
These fields will be integrated over in the path integral as
well, but their precise transformation properties need not
concern us if we simply include these fields in the $\phi^A$.
The explicit form of $\cal{R}_{Aa}$ in the case of an
irreducible closed algebra can be found in ref. \cite{anti}.

We can now straightforwardly combine these extended gauge
symmetries with the Schwinger-Dyson symmetries through the use
of collective fields. We again shift the fundamental fields
$\phi_A$, and we choose to include the internal gauge (BRST)
transformations in the transformation of the collective fields
only. The combined extended symmetries then take the form:
\begin{eqnarray}
\delta_a \phi_A & = & \phi^*_{Aa}  \cr
\delta_a \bar{\phi}_A &=&\phi^*_{Aa} -
{\cal R}_{Aa}^{(\phi-\bar{\phi})} \cr
\delta_a \phi^*_{Ab} & = & \epsilon_{ab} B_A \cr
\delta_a B_A & = & 0
\end{eqnarray}

Gauge fixing the collective fields to zero is again done by adding a
term manifestly invariant under the extended BRST symmetry,
\begin{eqnarray}
\frac{1}{4} \epsilon^{ab} \delta_a\delta_b
  [\bar{\phi}_A (-1)^{\epsilon_B}M_{AB} \bar{\phi}_B]
 & = & \epsilon^{ab} \phi^*_{Aa} M_{AB}
 {\cal R}_{Bb}^{(\phi - \bar{\phi})}  \cr
&  &- \frac{1}{2}(-1)^{\epsilon_B}\epsilon^{ab}
\bar{\phi}_A M_{AB} \frac{\delta^r
{\cal R}_{Bb}^{(\phi - \bar{\phi})}}{\delta \phi_C}
 {\cal R}_{C a}^{(\phi - \bar{\phi})} \cr
&  & + \bar{\phi}_A(-1)^{\epsilon_B+1} M_{AB} B_B -
\frac{1}{2} \epsilon^{ab}
\phi^*_{Aa} M_{AB} \phi^*_{Bb} \cr
&  &- \frac{1}{2} \epsilon^{ab}
{\cal R}_{Aa}^{(\phi - \bar{\phi})}
 M_{AB} {\cal R}_{Bb}^{(\phi-\bar{\phi})}
\end{eqnarray}
Notice that the ``antifields" $\phi^*_{Aa}$ (really the ghost-antighost
pair of the Schwinger-Dyson shift symmetries) automatically start
acting as sources for the internal BRST--anti-BRST symmetries,
while the collective fields $\bar{\phi}$ acts as a source for
the commutator of the two.

The extended action also contains terms quadratic in
the ghost-antighost fields $\phi^*_{Aa}$. Of course, when we integrate
over these fields (as well as over $\bar{\phi}_A$ and $B_A$), we
simply recover the starting action $S$ integrated over the standard
measure. The formulation in these terms is therefore entirely
consistent, and
we can now finally add a gauge-fixing term of the form
$\varepsilon^{ab}\delta_a\delta_b \chi(\phi)$. The interesting effect
of adding such a term is that the collective field $\bar{\phi}$ is
no longer set equal to zero in the gauge-fixing procedure. The
resulting gauge-fixed action seems at first sight to imply very
unusual Feynman rules since, $e.g.$, the argument of even the classical
action is
shifted by $\delta^r\chi(\phi)/\delta\phi_A$. It can, however,
formally be brought into the conventional form \cite{anti} by a
field redefinition. This aspect is avoided entirely if we choose
the ``gauge-boson" $\chi$ to be a function of the difference
$\phi_A - \bar{\phi}_A$
instead. The gauge-fixing is then directly identical to
the standard procedure. On the other hand, no new insight into
the gauge-fixing has been gained.

Instead, it is convenient to separate out the
terms of the
extended action that act as sources for the BRST--anti-BRST
transformations of the internal symmetries, and their commutator.
We therefore define one part $S_{BLT}$ of the extended action
$S_{ext}$ through
\begin{eqnarray}
S_{ext} & = & S_{0}(\phi - \bar{\phi}) + \frac{1}{4}\epsilon^{ab}
\delta_a\delta_b[\bar{\phi}_A(-1)^{\epsilon_B}M_{AB}
\bar{\phi}_B ] \cr
&=& S_{BLT} + \bar{\phi}_A(-1)^{\epsilon_B+1}M_{AB}B_B
-\frac{1}{2}\epsilon^{ab} \phi^*_{Aa}M_{AB} \phi^*_{Bb}
- \frac{1}{2} \epsilon^{a b}{\cal R}_{Aa}
M_{AB}{\cal R}_{Bb}   \cr
 & = & S_{BLT} + S_{aux}
\end{eqnarray}
Clearly, $S_{BLT}$ is nothing but the classical action plus the source
terms for the internal extended BRST transformations.
We have called what is left (the quadratic part of the ghost-antighost
fields $\phi^*_{Aa}$, the term that fixes the collective field
to zero, and the term independent of $\phi^*_{Aa}$
and $\bar{\phi}_A$) the ``auxiliary action", and denoted it by
$S_{aux}$. It is auxiliary in the sense that in the
end it only serves to remove the $\phi^*_{Aa}, \bar{\phi}_A$-fields
from the path integral in a correct manner. The reason for this
splitting-up of the extended action is twofold. First, it allows us
to formulate the construction of $S_{ext}$ by means of a
differential equation reminiscent of the conventional Master
Equations \cite{Batalin,us}. Second, it allows us to perform the
subsequent fixing of the original gauge symmetries in an elegant way,
uncovering a remnant of the gauge-fixing procedure of Batalin and
Vilkovisky \cite{Batalin},
At this final step we will make complete contact to the
previous work of Batalin, Lavrov and Tyutin \cite{Ext}.

Let us first discuss the Master Equations. At the level where the
collective fields and the auxiliary fields $B_A$ are still kept
in the action, these Master Equations take rather trivial forms.
They simply express invariance of the extended action $S_{ext}$
under both symmetries $\delta_a$. Thus, $0 = \delta_a S_{ext}$
immediately implies
\beq
\frac{\delta^r S_{ext}}{\delta\phi_A}\phi^*_{Aa} +
\frac{\delta^r S_{ext}}{\delta\bar{\phi_A}}(\phi^*_{Aa} -
\cal{R}_{Aa}^{(\phi-\bar{\phi})}) +
\frac{\delta^r S_{ext}}{\delta\phi^*_{Ab}}\varepsilon_{ab}B_A = 0~.
\eeq

But inserting the decomposition (16) into these equations give
us more conventional-looking Master Equations for the source
part $S_{BLT}$. (The easiest way to derive it is to note that
we are now, after gauge-fixing $\bar{\phi}_A$ to zero, free to
shift the fields $\phi_A \to \phi_A + \bar{\phi}_A$.) One finds
\beq
\frac{\delta^r S_{BLT}}{\delta\phi_A}{\cal R}_{Aa} +
\frac{\delta^r S_{BLT}}{\delta\bar{\phi}_A}\phi^*_{Aa} = 0~.
\eeq
Then, using the fact that $S_{BLT}$ contains the sources for
the internal BRST--anti-BRST transformations, we note that it can
be rewritten as
\beq
\frac{\delta^r S_{BLT}}{\delta\phi_A}\varepsilon_{ab}M_{AC}^{-1}
\frac{\delta^l S_{BLT}}{\delta\phi^*_{Cb}} + \frac{\delta^r
S_{BLT}}{\delta\bar{\phi}_A}\phi^*_{Aa} = 0~.
\eeq
After a linear field redefinition,
\beq
\phi^*_{Aa} = \varepsilon_{ba}\phi'^{*b}_B M_{BA}^{-1}~~,~~~~
\bar{\phi}_A = (-1)^{\varepsilon_B}\bar{\phi}'_BM_{BA}^{-1} ~,
\eeq
this yields
\beq
\frac{\delta^r S_{BLT}}{\delta\phi_A}\frac{\delta^l S_{BLT}}
{\delta\phi^{*a}_A} = \epsilon_{ab}\frac{\delta^l S_{BLT}}
{\delta\bar{\phi}_A}\phi^{*b}_A~.
\eeq
We have for convenience dropped primed indices on the new variables.
The l.h.s. of eq. (21) is nothing but half the generalized
antibracket $(S,S)_a$. Notice
that after the field-relabelling the matrix $M$ has disappeared
from $S_{BLT}$, which now takes the simple form
\beq
S_{BLT} = S_0(\phi) + \phi^{*a}_A\cal{R}_{Aa} + \frac{1}{2}
\bar{\phi}_A\varepsilon^{ab}\frac{\delta^r\cal{R}_{Aa}}
{\delta\phi_B}\cal{R}_{Bb}~.
\eeq

Due the redefinition of what we now view as ghost-antighosts
$\phi^{*a}_A$ and collective fields $\bar{\phi}_A$,
the ghost number assignments have changed. One easily sees that now
\beq
gh(\phi^{*a}_A) = (-1)^a - gh(\phi^A)~,~~~
gh(\bar{\phi}_A) = - gh(B_A) = -gh (\phi^A)~,
\eeq
while of course Grassmann parities remain unchanged.

Gauge-fixing in a manifest BRST--anti-BRST invariant manner is again
done by adding a term of the kind
\begin{eqnarray}
\frac{1}{2}\varepsilon^{ab}\delta_a\delta_b \chi(\phi)
&=&  \frac{1}{2} \varepsilon^{ab}\frac{\delta^r
\chi(\phi)}{\delta\phi_B}\frac{\delta^r\cal{R}_{Bb}}
{\delta \bar{\phi}_A}\cal{R}_{Aa} + \frac{1}{2}\varepsilon^{ab}
(-1)^{\varepsilon_B}\frac{\delta^r}{\delta \phi_A}
\frac{\delta^r}{\delta \phi_B} \chi(\phi) \cal{R}_{Aa}\cal{R}_{Bb}
\cr &=& - \frac{\delta^r\chi(\phi)}{\delta\phi_A}\frac{\delta^l
S_{BLT}}{\delta\bar{\phi}_A} + \frac{1}{2}\varepsilon^{ab}
(-1)^{\epsilon_B}\frac{\delta^r}{\delta\phi_A}\frac{\delta^r}
{\delta\phi_B}\chi(\phi)\frac{\delta^l S_{BLT}}{\delta
\phi^{*a}_A}\frac{\delta^l S_{BLT}}{\delta\phi^{*b}_B}
\end{eqnarray}

The partition function can then be written
\begin{eqnarray*}
{\cal{Z}} & = & \int [d\phi_A][d\phi^*_A] [d\bar{\phi}_A]
e^{\left[\frac{i}{\hbar} S_{aux} + S_{gf} + S_{BLT} \right]} \\
  & = & \int [d\phi_A][d\phi^*_A] [d\bar{\phi}_A] e^{\frac{i}{\hbar}
S_{aux}} \left[ \hat{U} e^{\frac{i}{\hbar} S_{BLT} } \right]~,
\end{eqnarray*}
with a ``gauge-fixing operator" defined by
\beq
\hat{U} \equiv \exp\left[-\frac{\delta^r\chi(\phi)}{\delta
\phi_A}\frac{\delta^l}{\delta\bar{\phi}_A}
- \frac{i\hbar}{2}\varepsilon^{ab}\frac{\delta^l}{\delta\phi_B}
\frac{\delta^r}{\delta\phi_A}\chi(\phi)\frac{\delta^l}{\delta
\phi^{*a}_A}\frac{\delta^l}{\delta\phi^{*b}_B} \right] ~.
\eeq

The Master equation (21) for $S_{BLT}$ corresponds to the one
proposed by Batalin,
Lavrov and Tyutin \cite{Ext}, and the gauge-fixing procedure through
the matrix $\hat{U}$ is also identical to theirs\footnote{In the
form given by Lavrov \cite{Lavrov}.}. The only difference
is that from the present derivation the ghost-antighost pair
$\phi^*_{Aa}$ are not set equal to zero after having operated with
$\hat{U}$ on $S_{BLT}$. Rather, these fields are now integrated over
with the ``measure"
\beq
d\mu[\phi^*] \equiv [d\phi^*]e^{\left[\frac{i}{\hbar}S_{aux}\right]}
{}~,
\eeq
but it is easy to see that the end result is the same. The
collective fields $\bar{\phi}_A$ {\em are} precisely set equal to
zero after the integration over $B_A$ in $S_{aux}$.
Through this derivation, we have thus
managed to stipulate in a very precise manner what is behind this
gauge-fixing procedure based on $\hat{U}$, and how the fields
$\phi^*_{Aa}, \bar{\phi}_A$ are to be treated in the path integral.
It is remarkable that the whole formalism originally could be
arrived at through algebraic considerations alone \cite{Ext},
apparently only guided by the principle of treating the ``antifields"
as sources for the internal BRST and anti-BRST symmetries (and their
independent combination in their commutator), and some analogies
with the Hamiltonian formalism \cite{Ham}.

We see now that the mysterious three sets of ``antifields"
$\phi^*_{Aa}$ and $\bar{\phi}_A$ are nothing but linear combinations
of the ghost-antighost pairs of ref. \cite{us}, and the collective
fields used to derive the Schwinger-Dyson BRST symmetry, respectively.
We now also understand the reason why these collective fields can not
immediately be
disposed of, as in the case of keeping just BRST (or anti-BRST)
symmetry: The lack of nilpotency of the operator $\varepsilon^{ab}
\delta_a\delta_b$ after having integrated out the collective fields
and their auxiliary fields ($B_A$). If one integrates
out {\em only} the collective fields, while keeping the $B_A$-fields,
nilpotency of the extended BRST operators may be kept. But the
resulting formalism does not appear very convenient.

As in the case of just BRST or anti-BRST symmetries, one sees that
the Master Equation must be satisfied due to much more general
principles than independence of the gauge-fixing functions. Also the
Master Equations of {\em extended} BRST symmetries ensure that correct
Schwinger-Dyson equations are obtained even after the subsequent
gauge-fixing of internal symmetries.

The fact that the partition function is independent of the
operator $\hat{U}$ follows in an entirely straightforward manner
from the present perspective: It simply reflects the addition of
an extended--BRST-exact term to the action. Since during the
introduction of this operator we have kept all fields, the
operator $\varepsilon^{ab}\delta_a\delta_b$ is nilpotent when
acting on an arbitrary function of all the fields $\phi_A,
\phi^*_{Aa}, \bar{\phi}_A$. The form of $\hat{U}$ derived here
is thus very specific for gauge bosons $\chi$ that only depend
on $\phi_A$. The general expression for $\varepsilon^{ab}
\delta_a\delta_b$ is straightforward to write out. Such a term
can be added to the action without changing the physical content,
but the resulting actions can look highly unusual. The addition
of such a general term $\varepsilon^{ab}\delta_a\delta_b$ to the
action (and the analogous modified $\hat{U}$-operation) has a
direct counterpart in the general operator
\beq
\hat{U} = e^{i\hbar\hat{T}(X)}~,~~~\hat{T} = \frac{1}{2}
\varepsilon_{ab}[\bar{\Delta}^b,[\bar{\Delta}^a,X]_-]_+
\eeq
which is discussed in refs. \cite{Ext,Lavrov}. Here $\bar{
\Delta^a} \equiv \Delta^a + (i/\hbar)\varepsilon^{ab}
\phi^*_{Ab}\delta^l/\delta\bar{\phi}_A$. The operator $\hat{T}$
is thus apparently the closest one can get to the analogue
of the ``quantum BRST--anti-BRST operator" equivalent of
$\varepsilon^{ab}\delta_a\delta_b$.

It should be very clear from our presentation that -- as in the
case of just BRST or anti-BRST symmetries --
there is no unique Lagrangian
quantization prescription. In the end we formulated the collective
field quantization in a manner that directly proved its equivalence
to the scheme of Batalin, Lavrov and Tyutin \cite{Ext}, but there
are obviously many ways of departing from that particular
formulation.

Since the Master Equation (21) has been {\em derived} from known
principles using the collective field technique, its present use
may seem rather limited. What is needed in order to go beyond the
derivation in terms of collective fields is an overall formulation
that ensures correct Schwinger-Dyson equations as Ward Identities,
without losing nilpotency of the operator $\varepsilon^{ab}
\delta_a\delta_b$. As long as the derivation is directly tied to
the collective fields, the answer in a sense needs to be known
beforehand. One can nevertheless formulate empirical rules and
boundary conditions for the Master Equations (21) that will
guarantee the correct prescription for the gauge-fixing. Whether
or not theories with open gauge algebras can be made invariant
under extended BRST symmetries (a problem that from the present
point of view requires that we go beyond the relatively
straightforward collective field technique used here), and how it
ties up with Master Equations related to those of eq. (21) remains
to be investigated.

Related to the close connection with collective fields is the
question of quantum corrections to the Master Equations. Such
terms of the kind $\Delta^a S$ do not appear when the derivation
can be phrased in terms of collective fields in the manner
described here, because then the functional measure is (formally)
guaranteed to be invariant. Ultraviolet regulators that break
some of the symmetries may correspond to non-invariant measures,
and then these terms should be included.

Although the Master Equations (21) can be written in terms of the
generalized antibracket, it is clear that the canonical structure
present when quantizing with either BRST or anti-BRST symmetry alone
is lost. This is entirely natural from the point of view of ref.
\cite{us}, since the full canonical structure only appears when
one integrates out one of the fields $\phi^*_{Aa}$, while keeping
its ghost or antighost partner in the path integral. In this
manifestly BRST--anti-BRST invariant formulation it cannot
be achieved, and the canonical formalism is therefore not fully
present. The same holds for the ``quantum BRST" (or anti-BRST)
operator, which is simply the Schwinger-Dyson BRST (anti-BRST)
operator after the integration of only one of the ghost-antighosts
$\phi^*_{Aa}$ \cite{us}. Such forms of the operators have no
natural r\^{o}les in our $Sp(2)$ invariant formulation either,
and the operator $\hat{T}$ above is probably the closest one can
get. It is, however, not a priori ruled out that the extended
BRST symmetries
can be introduced only based on the Schwinger-Dyson BRST operator.
Although $Sp(2)$ symmetry will not be manifest at all levels, it
may perhaps be imposed on the subspace of fields $\phi^A$, which,
after all, would be sufficient.

The geometric interpretation of the usual Batalin-Vilkovisky
quantization scheme has recently received considerable attention
\cite{Witten}. In an interesting paper, Henneaux \cite{Henneaux}
has shown that several aspects of this
geometric interpretation can be carried over to the case of
extended BRST symmetries when based on the
formulation given in ref. \cite{Ext}. An obvious open
problem is to give a geometric interpretation of the new
ghosts $c^A$ of ref. \cite{us}, without being restricted to
an $Sp(2)$-invariant formulation.

\end{document}